# When a covalent bond is broken?

Elena Sheka and Nadezhda Popova

Peoples' Friendship University of Russia,
117198 Moscow, Russia
sheka@icp.ac.ru

Speaking about the length of a covalent bond, one usually addresses the data tabulated in numerous tables and presented in numerous handbooks (see, for example, [1-3]). As seen from the data, bond lengths of the same order for the same pair of atoms in various molecules are very consistent which makes it possible to speak about standard values related to particular pairs of atoms. Thus, a standard length of 1.09Å is attributed to the C-H pair while the lengths of 1.54, 1.34, and 1.20Å are related to single, double and triple C-C bond, respectively. At the same time, the tabulated bond dissociation energy related to the C-H bond varies quite significantly depending on to which molecule the bond belongs. In case of series from methyl to t-butyl radicals the energy decreases from 103 to 93 kJ/mol, while taking values of 110, 85, 88, 87, and 112 kJ/mol in phenyl, benzyl, allyl, acetyl, and vinyl radicals, respectively. Changing in the bond length has been fixed therewith at the level of 1.09-1.05Å. Complicated as a whole, the set of the available data on bond lengths and bond energies provides a comprehensive view on the equilibrium state of molecules and solids.

In contrast, practically nothing is known when covalent bonds are broken. Usually one can find subjectively made estimations of critical values of the bond elongations that widely varied. The covalent bond breaking is considered as a final result of a continuous stretching. Such view explains why even weak loading greatly causes the acceleration of a body fragmentation [4]. However, a native question arises to which extent the bond can be stretched in order to be considered as a bond or, inversely, when it should be considered as broken? Recent success in experimental visualization of a chemical reaction using high-harmonic interferometry on the attosecond timescale [5] has not been able so far to fix the moment when the chemical bond is broken. On the other hand, the bond stretching is present in the chemical life not only in due course of mechanochemical reactions. Particular conditions of chemical reactions as well as peculiar properties of reactants may cause changing in the standard values of equilibrium chemical bond lengths thus making them stretched. This have been observed at the hydrogenation and fluorination of fullerene $C_{60}$ [6, 7], hydrogenation of graphene [8], and so forth. The C-H and C-F bond stretching in the last cases seems to exhibit a topochemical character of the relevant reactions, for which the knowledge concerning the high limit of the bond lengths is of a particular interest.

The current contribution suggests the application of the effectively-unpaired-electron concept to describe stretching and breaking of chemical bonds quantitatively. The approach, first suggested by Takatsuka, Fueno, Yamaguchi (TFY) over three decades ago [9], was elaborated by Staroverov and Davidson later on [10]. As shown, enlarging internuclear distances between valence electrons, which provide the covalent bond formation, causes the appearing of effectively unpaired electrons. The approach was firstly applied to the dissociation of $H_2$ and $O_2$ molecules [10] exhibiting the break of the relevant covalent bonds.

The radical character of a molecule is commonly perceived as a one-electron property. Although an open-shell singlet has arguably more radical character than a closed-shell species, the difference is not evident from conventional one-electron distributions. Indeed, the total charge density $\rho(r)$ by itself contains no implication of unpaired electrons, whereas the exact spin density $\rho_u(r) = \rho_\alpha(r) - \rho_\beta(r)$ for a singlet is zero at every position. To exhibit unpaired electrons, TFY suggested new density function

$$D(\mathbf{r}|\mathbf{r}') = 2\rho(\mathbf{r}|\mathbf{r}') - \int \rho(\mathbf{r}|\mathbf{r}'')\rho(\mathbf{r}''|\mathbf{r}')\,d\mathbf{r}'', \quad (1)$$

which characterizes the tendency of spin-up and spin-down electrons to occupy different portions of space. The function $D(r|r')$ was termed the distribution of 'odd' electrons, and its trace

$$N_D = \mathrm{tr}\, D(\mathbf{r}|\mathbf{r}'), \quad (2)$$

was interpreted as the total number of such electrons [9]. The authors suggested the function $D(r|r')$ trace $N_D$ to manifest the radical character of the species under investigation. Two decades later Staroverov and Davidson changed the term by the 'distribution of *effectively unpaired electrons*' [10, 11] emphasizing a measure of the radical character that is determined by the $N_D$ electrons taken out of the covalent bonding. Even in the TFY paper was mentioned [9] that the function $D(r|r')$ can be subjected to a population analysis within the framework of the Mulliken partitioning scheme. In the case of a single Slater determinant Eq. (2) takes the form [10]

$$N_D = tr DS, \quad (3)$$

where

$$DS = 2PS - (PS)^2. \quad (4)$$

Here $D$ is the spin density matrix $D = P^\alpha - P^\beta$, $P = P^\alpha + P^\beta$ is a standard density matrix in the atomic orbital basis, and $S$ is the orbital overlap matrix ($\alpha$ and $\beta$ mark different spin directions). The population of effectively unpaired electrons on atom $A$ is obtained by partitioning the diagonal of the matrix DS as

$$D_A = \sum_{\mu \in A}(DS)_{\mu\mu}, \quad (5)$$

so that

$$N_D = \sum_A D_A. \quad (6)$$

Staroverov and Davidson showed [11] that atomic population $D_A$ is close to the Mayer free valence index [12] $F_A$ in general case, while in the singlet state $D_A$ and $F_A$ are identical. Thus, a plot of $D_A$ over atoms gives a visual picture of the actual radical electrons distribution [13], which, in its turn, exhibits atoms with enhanced chemical reactivity.

In the framework of the unrestricted Hartree-Fock (UHF) approach, the effectively unpaired electron population is definitely connected with the spin contamination of the UHF solution state caused by single-determinant wave functions which results in a straight relation between $N_D$ and square spin $\langle \hat{S}^2 \rangle$ [11]

$$N_D = 2\left(\langle \hat{S}^2 \rangle - \frac{(N^\alpha - N^\beta)^2}{4}\right), \quad (7)$$

where

$$\langle \hat{S}^2 \rangle = \left(\frac{(N^\alpha - N^\beta)^2}{4}\right) + \frac{N^\alpha + N^\beta}{2} - \sum_i^{N^\alpha}\sum_j^{N^\beta}|\langle \phi_i | \phi_j \rangle|^2 \quad (8)$$

Here $\phi_i$ and $\phi_j$ are atomic orbitals; $N^\alpha$ and $N^\beta$ are the numbers of electrons with spin $\alpha$ and $\beta$, respectively.

If UHF computations are realized via *NDDO* approximation (the basis for AM1/PM3 semiempirical techniques) [14], a zero overlap of orbitals in Eq. (8) leads to $S = I$, where $I$ is the identity matrix. The spin density matrix $D$ assumes the form

$$D = (P^\alpha - P^\beta)^2. \quad (9)$$

The elements of the density matrices $P_{ij}^{\alpha(\beta)}$ can be written in terms of eigenvectors of the UHF solution $C_{ik}$

$$P_{ij}^{\alpha(\beta)} = \sum_{k}^{N^{\alpha(\beta)}} C_{ik}^{\alpha(\beta)} * C_{jk}^{\alpha(\beta)} . \qquad (10)$$

Expression for $\langle \hat{S}^2 \rangle$ has the form [15]

$$\langle \hat{S}^2 \rangle = \left( \frac{(N^\alpha - N^\beta)^2}{4} \right) + \frac{N^\alpha + N^\beta}{2} - \sum_{i,j=1}^{NORBS} P_{ij}^\alpha P_{ij}^\beta \qquad (11)$$

This explicit expression is the consequence of the $\Psi$-based character of the UHF approach. Since the corresponding coordinate wave functions are subordinated to the definite permutation symmetry, each value of spin $S$ corresponds to a definite expectation value of energy [16]. Oppositely, the electron density $\rho$ is invariant to the permutation symmetry. The latter causes a serious spin multiplicity problem for the UDFT schemes [17]. Additionally, the UDFT spin density $D(r|r')$ depends on spin-dependent exchange and correlation functionals only and cannot be expressed analytically [16]. Since the exchange-correlation composition deviates from one method to the other, the spin density is not fixed and deviates alongside with the composition.

Within the framework of the *NDDO* approach, the total $N_D$ and atomic $N_{DA}=D_A$ populations of effectively unpaired electrons take the form [18]

$$N_D = \sum_A N_{DA} = \sum_{i,j=1}^{NORBS} D_{ij} \qquad (12)$$

and

$$N_{DA} = \sum_{i \in A} \sum_{B=1}^{NAT} \sum_{j \in B} D_{ij} . \qquad (13)$$

Here $D_{ij}$ present matrix elements of the spin density matrix $D$.

Firstly applied to fullerenes [18-20], $N_{DA}$ in the form of Eq. (13) has actually disclosed the different chemical activity of atoms just visualizing the 'chemical portrait' of the molecule. It was naturally to rename $N_{DA}$ as *atomic chemical susceptibility* (ACS). Similarly referred to, $N_D$ was termed as *molecular chemical susceptibility* (MCS). Rigorously computed ACS ($N_{DA}$) is an obvious quantifier that highlights targets to be the most favorable for addition reactions of any type at each stage of the reaction thus forming grounds for *computational chemical synthesis*. A high potentiality of the approach will be exemplified by fluorination [6] and hydrogenation (7) of fullerene $C_{60}$ as well as hydrogenation of graphene [8]. An accumulating review is presented in [21].

Oppositely to UHF, UDFT does not suggest enough reliable expressions for either $N_D$ or $N_{DA}$. The only known detailed discussion of the problem comparing UHF and UDFT results with complete active space selfconsistent field (CASSCF) and multireference configuration interaction (MRCI) ones concerns the description of diradical character of the Cope rearrangement transition state [22]. When UDFT calculations gave $N_D = 0$, CASSCF, MRCI, and UHF calculations gave 1.05, 1.55, and 1.45 $e$, respectively. Therefore, experimentally recognized radical character of the transition state was well supported by the latter three techniques with quite a small deviation in numerical quantities while UDFT results just rejected the radical character of the state. Serious UDFT problems are known as well in the relevance to $\langle \hat{S}^2 \rangle$ calculations [23, 24]. These obvious shortcomings of the UDFT approach might be a reason why UDFT calculations of this kind are rather scarce.

Analysis of the $N_D$ values behavior along the potential energy curve of diatomic molecules firstly performed by Staroverov and Davidson [10] and then repeated by Sheka and Chernozatonskii in [25] has given an excellent possibility to check the correctness of the UHF approach to the description of effectively unpaired electrons. As shown, a characteristic *S*-like character of the $N_D(R)$ dependence is common for all molecules. Each *S*-curve involves three regions, namely, (I) $R \leq R_{cov}$, (II) $R_{cov} \leq R \leq R_{rad}$, and (III) $R \geq R_{rad}$. At $R<R_{cov}$, $N_D(R)=0$ and $R_{cov}$ marks the extreme distance that corresponds to the completion of the covalent bonding of the molecule electrons and which exceeding indicates the onset of the molecules radicalization. $R_{rad}$ matches a completion of homolytic bond cleavage followed by the formation of two free radicals with practically constant value of $N_D(R) = N_D^{rad}$. The intermediate region II with a continuously growing $N_D$ value from zero to $N_D^{rad}$ exhibits a continuous build-up of the molecular fragments radicalization caused by electron extraction from the covalent bonding as the corresponding interatomic bond is being gradually stretched.

It seems quite reasonable that similar S-like curve should be expected for any chemical bonds. To check the assumption, we have performed UHF calculations of a series of $N_D(R)$ curves related to C-C bonds of different order and to C-H bond in ethane. Obtained results are presented in Figure 1. The computations have been carried out by using both AM1 and PM3 semiempirical versions of CLUSTER-Z1 codes [26]. Based on the *NDDO* approximation, the codes are very efficient, accurate, and suitably enlarged by including calculation of square spin $\langle \mathcal{S}^2 \rangle$ (Eq. (11)) as well as effectively unpaired electrons population analysis in terms of $N_D$ (Eq.(12)) and $N_{DA}$ (Eq. (13)) [18]. A comprehend description of the codes is given elsewhere [27] alongside with its parallel version. The bond elongation was performed in a stepwise manner with increments of 0.05Å in general and of 0.02Å when some details were considered more scrupulously.

*C-C bonds*. We have chosen three molecules (ethane, cyclohexane and hexamethylcyclohexane) to analyze single C-C bonds and other three molecules (ethylene, benzene and hexamethylbenzene) to investigate the $N_D(R)$ curves for double C-C bonds. As seen in the figure, all the studied $N_D(R)$ curves are of S-like shape, quite similar within each of two sets, but significantly different between the sets: the single-C-C-bond $N_D(R)$ curves are of a one-stage S-shape while for double C-C bonds S-like curves are of two stages. As seen in Figure 1, S-curves for single C-C bonds are characterized by abrupt transitions within the region of $R_{cov}$ and $R_{rad}$ points which allows a strict fixation of both values. In contrast, while the fixation of $R_{cov}$ points on the S-curves of double C-C bonds is well designed, the position of the $R_{rad}$ points is quite uncertain due to extended weakly increased ranges on the $N_D(R)$ curves in the final part of the curves. The data concerning $R_{cov}$ and $R_{rad}$ values alongside with equilibrium bond lengths $l_{eq}$ and $N_D^{rad}$ are listed in Table 1.

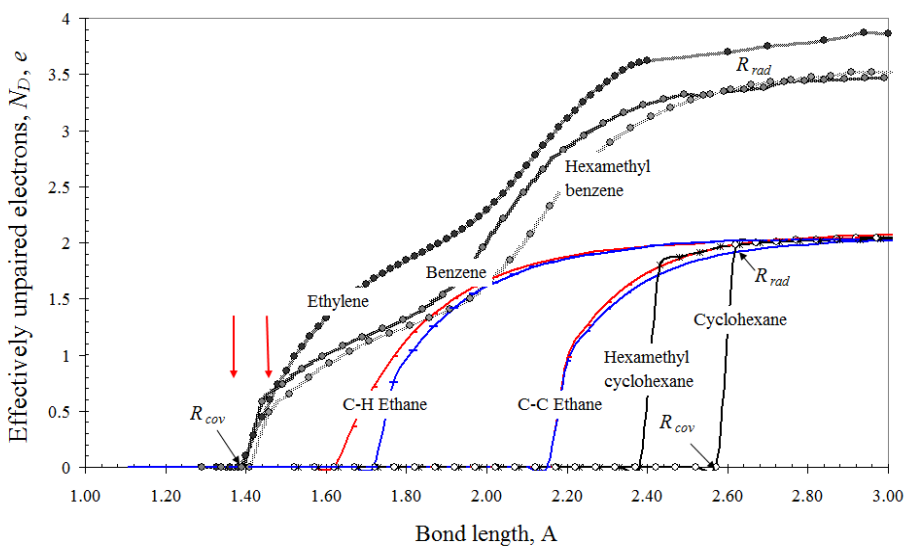

**Figure 1**. The number of effectively unpaired electrons in a set of molecules as a function of the internuclear C-C separation (black curves). Red and blue curves are related to C-H and C-C separations of the ethane molecule computed by application of AM1 and PM3 versions of the CLUSTER-Z1 codes, respectively.

Table 1. Internuclear separations and the number of effectively unpaired electrons under bond's break

| Molecule | $l_{eq}$, Å | $R_{cov}$, Å | $R_{rad}$, Å | $N_D^{rad}$, e |
|---|---|---|---|---|
| C-C bonds | | | | |
| ethane | 1.50 | 2.15 | 2.65 | 2.01 |
| cyclohexane | 1.52 | 2.57 | 2.62 | 2.02 |
| hexamethylcyclohexane | 1.53 | 2.38 | 2.43 | 2.00 |
| ethylene | 1.32 | 1.40 | 1.90/2.70 | 3.86 |
| benzene | 1.39 | 1.39 | 1.99/2.59 | 3.47 |
| hexamethylbenzene | 1.41 | 1.41 | 1.96/2.61 | 3.52 |
| C-H bonds | | | | |
| ethane | 1.117 | 1.617/1.717[1)] | 2.30 | 1.98 |

[1)] The data are related to AM1 and PM3 calculations, respectively.

As seen from the table, single C-C bonds are subjected to a considerable stretching first stage of which is determined by $R_{cov}$. The stretching constitutes from 43 to 69% of the initial equilibrium bond length and further its elongation results in achieving $R_{rad}$ quite quickly. In contrast to the case, $R_{cov}$ on the $N_D(R)$ curves related to the double C-C bonds exceeds $l_{eq}$ only for ethylene molecule while for benzene and hexamethylbenzene $R_{cov}$ and $l_{eq}$ coincide. A further stretching over $R_{cov}$ opens the bond radicalization, the first stage of which is marked by approaching $N_D$ to 2 (clearly seen kinks on $N_D(R)$ curves in the region of 1.90-2.00Å). Then follows the second stage that is completed by a full radicalization of two molecular fragments formed at 2.60-2.70 Å. Therefore, in spite of a drastic difference in the behavior of the $N_D(R)$ curves related to either single or double C-C bonds, a full radicalization of the formed fragments occurs at 2.60-2.70Å in both cases. Answering the question put in the paper title, one should state that the covalent C-C bond is broken when its length achieves $R_{rad}$. The value is practically the same for both single and double bonds and is located within the interval of 2.60-2.70Å. The bond break is followed by two prior-breaking periods of the bond elongation. The first period locates the interval between $l_{eq}$ and $R_{cov}$, where the bond elongation, quite considerable for single bonds and small, if observable at all, for double bonds, does not change the bond character. The second period commences at $R_{cov}$ and concerns a gradual radicalization of the molecular fragments formed. The width of the period is quite short in the case of single bonds and, oppositely, quite large for double (and triple) covalent bonds. The feature is one of the most vivid exhibitions of the odd-electron character of the double C-C bonds that lays the foundation of all the peculiarities of the electron behavior in $sp^2$ nanocarbons [21]. Two vertical arrows in Figure 1 mark the C-C bond length interval characteristic for both fullerene $C_{60}$ and carbon nanotubes and graphene. In all the cases the interval is located in the intermediate region above $R_{cov}$.

*C-H bonds*. The bonds behavior is similar to that of single C-C bonds. As seen in Figure 1, the $N_D(R)$ curve behaves quite similarly to the C-C bond of the same molecule, but is shifted to shorter bond length region due to smaller value of the equilibrium length $l_{eq}$. As in the case of C-C bonds, C-H ones can be considerably stretched without changing the bond character, achieving 45% elongation and characterized by $l_{max} \cong 2.30$Å.

Summing up the discussion of results related to gradual elongation of C-C and C-H bonds, one can conclude the following:

The maximum values of the bond length $l_{max}$ are about twice larger than equilibrium ones $l_{eq}$ for bonds of both types.

Due to the similarity of the H-H,

N-N, O-O covalent bonds behavior [10, 25] to that of C-C and C-H ones, a similar ratio $l_{max}/l_{eq}$ should be expected for any kind of covalent bonds.

The finding should be taken into account when inputting structural data in programs of graphic presentation of molecular structures.

References

1. R.T.Sanderson, *Polar Covalence*, 1983
2. R.T.Sanderson, *Chemical Bonds and Bond Energy*, 1976
3. B. deB. Darwent, "National Standard Reference Data Series," National Bureau of Standards, No. 31, Washington, DC, 1970
4. T. J. Kucharski, Z. Huang, Q.-Z. Yang, Y. Tian, N. C. Rubin, C. D. Concepcion, R. Boulatov. Kinetics of thiol/disulfide exchange correlate weakly with the restoring force in the disulfide moiety. *Ang. Chem. Int. Ed*. 121 (2009) 7057.
5. H. J. Wörner, J. B. Bertrand, D. V. Kartashov, P. B. Corkum, and D. M. Villeneuve. Following a chemical reaction using high-harmonic interferometry. *Nature*, July 28 (2010) 604.
6. E. F. Sheka. Stepwise computational synthesis of fullerene $C_{60}$ derivatives. Fluorinated fullerenes $C_{60}F_{2k}$. *J. Exp. Theor. Phys*. 111(2010) 395.
7. E. F. Sheka. Computational synthesis of hydrogenated fullerenes from $C_{60}$ to $C_{60}H_{60}$. *J. Mol. Mod*. 17 (2011) 1973.
8. E. F. Sheka and N.A.Popova. How graphene is transformed into regular graphane structure. arXiv:1108.3979v1 [cond-mat.mtrl-sci].
9. K.Takatsuka, T.Fueno, and K.Yamaguchi. Distribution of odd electrons in ground-state molecules. *Theor. Chim. Acta* 48 (1978) 175.
10. V. N. Staroverov and E. R. Davidson. Distribution of effectively unpaired electrons. *Chem. Phys. Lett*. 330 (2000) 161.
11. V.N. Staroverov and E.R. Davidson. Electron distribution in radicals. *Int. J. Quant. Chem*. 77 (2000) 316.
12. I. Mayer, I. On bond orders and valences in the *ab initio* quantum chemical theory. *Int. J. Quant. Chem*. 29 (1986) 73.
13. E.R. Davidson and A.E.Clark. Analysis of wave functions for open-shell molecules. *Phys. Chem. Chem. Phys*. 9 (2007) 1881.
14. M.J.S.Dewar, and W.Thiel, Ground states of molecules. 38. The MNDO method. Approximations and parameters. *J. Am. Chem. Soc.* 99 (1977) 4899.
15. D.A.Zhogolev, and V.B.Volkov. *Methods, Algorithms and Programs for Quantum-Chemical Calculations of Molecules* (in Russian). Kiev: Naukova Dumka, 1976.
16. I.Kaplan. Problems in DFT with the total spin and degenerate states. *Int. J. Quant. Chem*. 107 (2007) 2595.
17. E. Davidson. How robust is present-day DFT? *Int. J. Quant. Chem*. 69 (1998) 214.
18. E.F.Sheka, and V.A.Zayets. The radical nature of fullerene and its chemical activity. *Russ. J. Phys. Chem*. 79 (2005) 2009.
19. E.F.Sheka. Chemical portrait of fullerenes. *J. Struct. Chem*. 47 (2006) 593.
20. E.F.Sheka. Chemical susceptibility of fullerenes in view of Hartree-Fock approach. *Int. J. Quant. Chem*. 107 (2007) 2803.
21. E.F.Sheka. *Fullerenes: Nanochemistry, Nanomagnetism, Nanomedicine, Nanophotonics.* CRC Press, Taylor and Francis Group: Boca Raton, 2011.
22. V.N. Staroverov and E.R. Davidson. Diradical character of the Cope rearrangement transition state. *J. Am. Chem. Soc*. 122 (2000) 186.
23. J.Wang, A.D. Becke, and V.H.Smith, Jr. Eveluation of $\langle S \rangle^2$ in restricted, unrestricted Hartree-Fock, and

density functional based theory. *J. Chem. Phys*. 102 (1995) 3477.
24. A.J.Cohen, D.J. Tozer, and N.C.Handy. Evaluation of $\langle S \rangle^2$ in density functional theory. *J. Chem. Phys*. 126 (2007) 214104 (4pp).
25. E.F.Sheka and L.A.Chernozatonskii. Bond length effect on odd electrons behavior in single-walled carbon nanotubes. *J. Phys. Chem. C* 111 (2007) 10771.
26. V.A.Zayets. *CLUSTER-Z1: Quantum-Chemical Software for Calculations in the s,p-Basis*. Kiev: Institute of Surface Chemistry Nat. Ac.Sci. of Ukraine, 1990.
27. P.K. Berzigiyarov, V.A. Zayets, I.Ya. Ginzburg, V.F. Razumov, and E.F.Sheka. NANOPACK: Parallel codes for semiempirical quantum chemical calculations of large systems in the *sp*- and *spd*-basis. *Int. J. Quant. Chem.* 88 (2002) 449.
28. J.Hachmann, J.J. Dorando, M. Aviles, and G.K.Chan. The radical character of the acenes: A density matrix renormalization group study. *J. Chem. Phys*. 127 (2007) 134309 (9pp).

1.